# OPTIMIZATION OF HIGHLY CIRCULARLY POLARIZED THERMAL RADIATION IN α-MoO$_3$/β-Ga$_2$O$_3$ TWISTED LAYERS

Marco Centini[1*], Chiyu Yang[2], Maria Cristina Larciprete[1], Mauro Antezza[3,4], Zhuomin M. Zhang[2]

[1]Sapienza University of Rome Department of Basic and Applied Sciences for Engineering,
Via A. Scarpa 14, I-00161 Rome, Italy
[2]George W. Woodruff School of Mechanical Engineering,
Georgia Institute of Technology, Atlanta, GA 30332, USA
[3]Laboratoire Charles Coulomb (L2C), UMR 5221 CNRS-Université de Montpellier,
F- 34095 Montpellier, France
[4]Institut Universitaire de France, 1 rue Descartes, F-75231 Paris Cedex 05, France

ABSTRACT. We investigate a bi-layer scheme for circularly polarized infrared thermal radiation. Our approach takes advantage of the strong anisotropy of low-symmetry materials such as β-Ga$_2$O$_3$ and α-MoO$_3$. We numerically report narrow-band, high degree of circular polarization (over 0.85), thermal radiation at two typical emission frequencies related to the excitation of β-Ga$_2$O$_3$ optical phonons. Optimization of the degree of circular polarization is achieved by a proper relative tilt of the crystal axes between the two layers. Our simple but effective scheme could set the basis for a new class of lithography-free thermal sources for IR bio-sensing.

## 1. INTRODUCTION

The development of miniaturized and integrable devices operating in the mid-infrared (mid-IR) relies on the availability of narrowband, highly directional, and polarization tunable sources. Indeed, several applications ranging from IR sensing [1,2] to biomedical diagnostics and target detecting [3,4] as well as for the realization of integrated IR photonics [5,6] require highly efficient, narrow-band, and polarization tunable mid-IR sources. An efficient but expensive option is represented by quantum cascade laser (QCL) IR sources. However, with the aim to achieve mass production of low-cost devices, a lot of research effort has been recently spent on the study of optimized and controllable thermal radiation sources. Thanks to the development of micro/nano technologies [7], several approaches to control and tailor the typically incoherent and broadband thermal radiation from heated bodies have been proposed and realized. The mechanism at the base of this tailored emission is the excitation of resonant states i.e. surface plasmon-(SP) or phonon-(SPh) polaritons, Bloch surface waves, magnetic polaritons, to name a few. Gratings [8], metamaterials [9,10], metasurfaces [11] and nanoantennae [12,13] have been successfully proposed to overcome and enhance the performances of natural bulk materials. However, all these approaches rely on the use of lithographic techniques to artificially create absorption/emission resonances as well as a strong birefringent effective response. In particular, birefringence is necessary for polarization sensitive applications.

An extreme form of birefringent behavior is called hyperbolicity and it is achieved when one diagonal element of the material dielectric tensor is negative, whilst the others are positive [14]. Hyperbolic metamaterials obtained by periodic sub-wavelength patterned layered media have been studied for enhanced thermal radiation and nanoscale heat transfer [15,16]. However, it has been recently shown that hyperbolicity can be observed in natural materials, more specifically in van der Waals (vdW) materials IR [17]. Among vdW materials, hexagonal boron nitride hBN is one of the most investigated [18]. The hBN exhibits two different Restsrahlen bands (negative values of the real part of the permittivity) along either the in-plane or out-of-plane directions. The main limitation of hBN is the lack of in-plane anisotropy. In-plane anisotropy is required to distinguish and manipulate orthogonal polarization states of the electromagnetic field at normal incidence.

---

[*] Corresponding Author: marco.centini@uniroma1.it

Among the vdW polar materials displaying in-plane anisotropy, molybdenum trioxide (α-$MoO_3$) has been the object of a conspicuous number of recent studies [19,20]. Its strong anisotropy, allowing in-plane and off-plane hyperbolicity, has been used to obtain mid-IR waveplates [21] and tunable [22] and broadband [23] absorption in the far-field. Concerning the excitation of surface waves, hyperbolic phonon polaritons have recently been shown in α-$MoO_3$ flakes from 818 cm$^{-1}$ to 974 cm$^{-1}$ [20,24]. Furthermore, the excitation of SPh polaritons waves in twisted flakes of 2D α-$MoO_3$ has been proposed and experimentally verified [25]. Such control of SPh polaritons in twisted flakes has interesting applications for tunable near-field radiative heat transfer [26] by adjusting the tilt angle between the receiver and the emitter. It has been shown that twisted α-$MoO_3$ layers can exhibit circular dichroism [27,28] and can be used to generate spin thermal radiation [29]. However, using twisted structures of identical anisotropic materials to achieve spin thermal radiation has more restrictions on the materials. For this reason, α-$MoO_3$ was combined with an ideal quarter wavelength plate [29]. However, a tunable combination of two real anisotropic materials (i.e. controlling the tilt angle between them) could be adopted for practical applications and for integrated thermal sources.

Low-symmetry materials have recently emerged as possible candidates for anisotropic optical applications [30]. Among them, β-$Ga_2O_3$ [31,32] has been used to experimentally demonstrate the excitation of shear phonon polaritons in the IR [33,34]. Moreover, this material perfectly matches the usable wavelength range of α-$MoO_3$. More specifically we will show that there are phonon resonances in the β-$Ga_2O_3$ which fall in an almost transparent but highly anisotropic band of the α-$MoO_3$. Thus, we propose a combination of two twisted layers obtained from these two materials.

Our calculations, based on a complete Stokes parameters analysis [35] of emitted radiation, show that it is possible to obtain circularly polarized thermal radiation in the mid-IR with a twisted bilayer system having about two microns of total thickness. As a starting point we study the emission properties and the birefringence behavior of single layers of α-$MoO_3$ and β-$Ga_2O_3$ on gold substrates. The choice of gold substrate allows for an almost one-sided emissivity as long as we can ignore thermal emission for the gold back face in the mid-IR. After the choice of one of the two materials as the emitter, we add a properly tilted and sized layer of the other, acting as a quarter wave plate to obtain thermally radiated circularly polarized light.

## 2. NUMERICAL METHODS

In order to quantitatively study the degree of polarization (DoP) and the degree of circular polarization (DoCP) of the Thermal radiation polarization properties can be characterized by Stokes parameters. For reciprocal materials, the Stokes parameters can be expressed in terms of the polarized emissivities as [35,36]:

$$\begin{bmatrix} S_0 \\ S_1 \\ S_2 \\ S_3 \end{bmatrix} = \frac{S_{0,bb}(\omega,k_\parallel)}{2} \begin{bmatrix} \epsilon_p + \epsilon_s \\ \epsilon_p - \epsilon_s \\ \epsilon_{45°} - \epsilon_{135°} \\ \epsilon_R - \epsilon_L \end{bmatrix}; \qquad (1)$$

where $k_\parallel$ is the wave vector projection onto the surface (x-y) plane, $S_{0,bb}$ is the emissivity of the black body, $\epsilon_{p,s}$ are the *p*- and *s*- polarized relative emissivities, $\epsilon_{45°,135°}$ are the 45°- and 135°- polarized relative emissivities and $\epsilon_{R,L}$ are the *right*- and *left*- circular polarized relative emissivities respectively. Limiting our discussion to reciprocal media at thermal equilibrium, polarized emissivities can be evaluated by calculating the polarized absorptivities α, according to Kirchhoff's law:

$$\epsilon_{p,s,45°,135°,R,L} = \alpha_{p,s,45°,135°,R,L} \qquad (2)$$

Polarized absorptivities have been retrieved by evaluating 1–*R*–*T* where *R* is reflectance and *T* is transmittance evaluated for the different polarized states of the incident light with a 4×4 transfer matrix method [37]. The degree of polarization (DoP) and the degree of circular polarization (DoCP) are then defined as:

$$\text{DoP} = \frac{\sqrt{S_1^2 + S_2^2 + S_3^2}}{S_0}; \qquad (3)$$

$$\text{DoCP} = \frac{|S_3|}{S_0}; \qquad (4)$$

Being both DoP and DoCP limited in the range [0,1] and DoCP≤DoP. DoP=1 stands for perfectly polarized light, thus DoCP=1 corresponds to perfectly circularly polarized light.

Our approach is based on the combination of two layers on top of gold substrate. The bottom layer acts as an emitter while the second acts as a quarter-wave plate. Both layers are required to be strongly anisotropic, however, the emitting layer should provide high emission/absorption efficiency for only one polarization component while the upper one should be as transparent as possible with a strong anisotropic real part of the refractive index. A particle swarm optimization is used as the search algorithm for the optimal parameters including the geometrical parameters (thickness) and the arrangement of the optic axes with the relative tilt angle between the two layers. In order to reduce the number of optimization parameters for the two-layer system the particle swarm algorithm is performed in two separate steps. At first, we set as the objective function the maximum degree of polarization (DoP) of the single, bottom, layer on gold. This way we find the layer thickness and emission wavelength for maximized DOP. Then we optimize the thickness and the relative orientation of the second (upper) layer to obtain maximum DoCP. We focus on the fact that this two-step optimization is performed to reduce the number of simultaneous sweeping parameters but it does not affect the accuracy of numerical results. Indeed, the adopted 4x4 transfer matrix model takes into account for multiple reflections. A complete single-step full sweep could lead to improved performances at a slightly shifted emission line. Nevertheless, we expect that this shift, typically related to multiple reflections, could play a relevant role in multilayer stacks or high-Q cavities, here we only focus on the double-layer scheme.

## 3. MATERIALS OPTICAL PROPERTIES

We study both α-MoO₃ and β-Ga₂O₃ optical properties in the mid-IR with the aim of identifying the frequency ranges where they can be efficiently combined. To begin with, we focus our study on the thermal radiation emitted along the direction normal to the surface. Thus, if we consider the *x-y* plane as the surface plane, the emission direction is toward the *z*-axis. We also select the orientation of the crystals so that *x*- and *y*- axes coincide with the *x'*- and *y'*- crystal axes. In this case, we are interested in the in-plane values of the dielectric permittivities. In Figure 1(a,b), the imaginary part of *xx, yy* and *xy* components of the relative permittivity tensor of α-MoO$_3$ and β-Ga$_2$O$_3$ are plotted as a function of the frequency [20,38]. The imaginary part of the dielectric function is related to absorption/emission. For α-MoO$_3$, there are two main peaks at 550 cm$^{-1}$ and 820 cm$^{-1}$ due to optical phonons while for β-Ga$_2$O$_3$, there are three main bands centered around 570 cm$^{-1}$, 690 cm$^{-1}$ and 740 cm$^{-1}$, respectively. As previously mentioned, in our emitter/waveplate approach we want to avoid overlaps in the emission bands in order to clearly distinguish the role of the emitter from the role of the waveplate. For this reason, we exclude from further investigation the lower frequencies (570 cm$^{-1}$ and 550 cm$^{-1}$). In Figure 1(c), the real part of the refractive index components $n_x$ and $n_y$ α-MoO$_3$ and their $\Delta n = n_x - n_y$ are shown. The real part of the dielectric function components of β-Ga$_2$O$_3$ are shown in Figure 1(d). We note that α-MoO$_3$ has a wide band (from 600 cm$^{-1}$ to 750 cm$^{-1}$) of high and almost constant anisotropy which makes it very suitable to be used as a waveplate. Thus we focus our attention on the design of an efficient β-Ga$_2$O$_3$ emitting layer around 690 cm$^{-1}$ and 740 cm$^{-1}$ with a high DoP combined with a properly tilted and sized α-MoO$_3$ layer acting as waveplate to obtain a tailored thermal source with optimized DoCP.

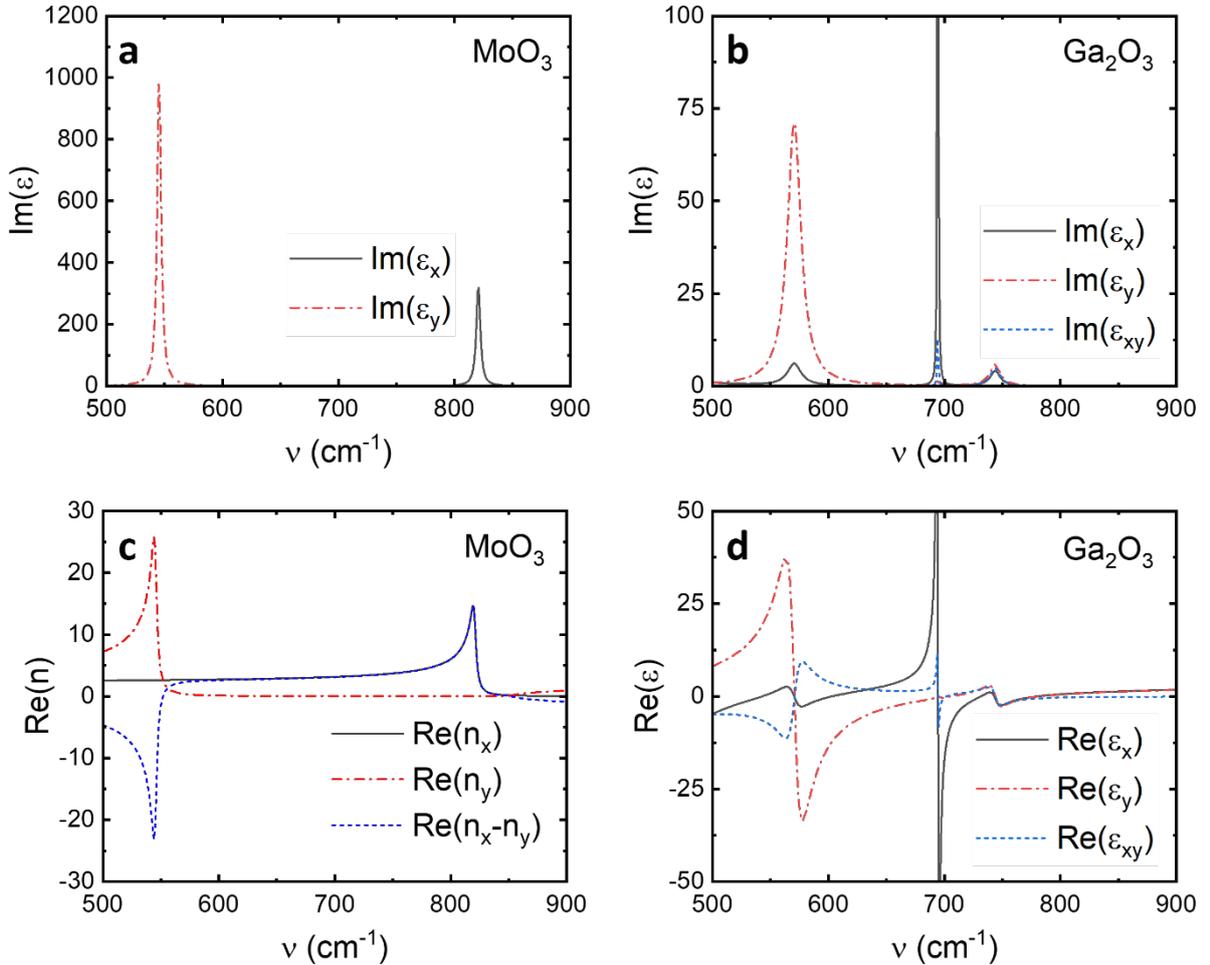

Figure 1. Imaginary part of the in-plane components (xx,xy,yy) of the dielectric permittivity tensor of (a) MoO$_3$, (b) Ga$_2$O$_3$. Real part of the refractive indices along x ($n_x$) and y ($n_y$) directions and Δn for (c) MoO$_3$ and Real part of the relative permittivity for (d) Ga$_2$O$_3$.

## 4. RESULTS AND DISCUSSION

We employed particle swarm optimization (PSO), a search algorithm that finds optimal solutions by iteratively trying to improve candidate solutions, and finally obtained structures with the highest circular emission at specific frequencies. The first optimization is performed on a single β-$Ga_2O_3$ layer on gold substrate. We found two schemes corresponding to maximized DoP at two different frequencies: 691 $cm^{-1}$ and 737 $cm^{-1}$. In the first scheme, sketched in Figure 2a, we note that a 606 nm thick β-$Ga_2O_3$ layer has a maximum relative emissivity for a p-polarized field at 691 cm-1 corresponding to a DoP of about 0.97 and zero DoCP (Figure 2b). On the other hand, the second proposed scheme, sketched in Figure 2c is based on a thicker Ga2O3 layer (1909 nm) showing a maximum DoP=0.93 at ν=737 $cm^{-1}$. The main difference with respect to the previous case is that the initial DoCP is now about 0.1 (figure 2d). Thus the emitted radiation from the single Ga2O3 layer is elliptical.

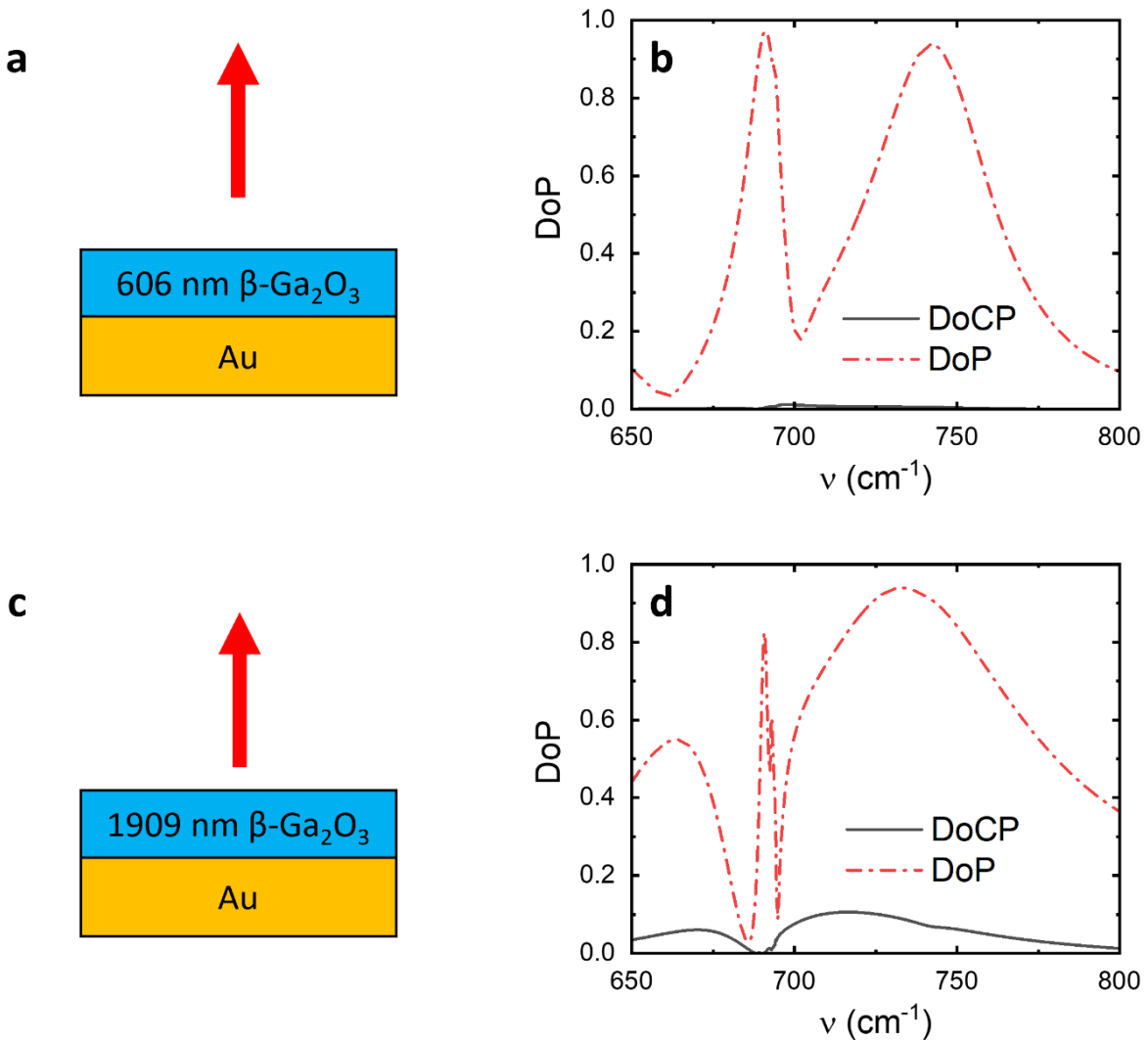

Figure 2. (a) Schematic of a single β-$Ga_2O_3$ layer exhibiting optimized DoP at ν=691 $cm^{-1}$; (b) DoP and DoCP as a function of the frequency for the system sketched in (a); (c) Schematic of a single β-$Ga_2O_3$ layer exhibiting optimized DoP at ν=737 $cm^{-1}$; (d) DoP and DoCP as a function of the frequency for the system sketched in (c).

Then we added a MoO3 cap layer to both schemes and applied a second PSO varying the thickness and the relative orientation with respect to the bottom layer. For the first scheme we found an optimal condition corresponding to a DoCP=0.87 at 691 cm$^{-1}$ when a 740 nm α-MoO$_3$ layer, tilted by an angle of 83° in the *x-y* plane with respect to the *x'y'* axes of the β-Ga$_2$O$_3$ layer is considered (sketch of figure 3a). All the results have been obtained with the indirect method, (i.e., calculating the absorptivity and applying Kirchhoff's law) and validated by the direct method based on fluctuation electrodynamics [35]. Figure 3b shows a high peak of almost left circularly polarized (LCP) emission at ν=691 cm$^{-1}$. The corresponding DoP and DoCP are reported in Figure 3c. We note that DoP=DoCP=0.87 at nearly the same frequency (see Figure 3c). Finally, we report in Figure 3d the value of S3/S0 as a function of the twist angle between the two layers and of the frequency. Its modulus is the DOCP, however, its sign determines the Right (if positive) or Left (if negative) circular polarization. The reported map clearly shows that at ν=691 cm$^{-1}$ it is possible to switch from almost perfect (DoCP=0.87) RCP to LCP by properly tuning the twist angle.

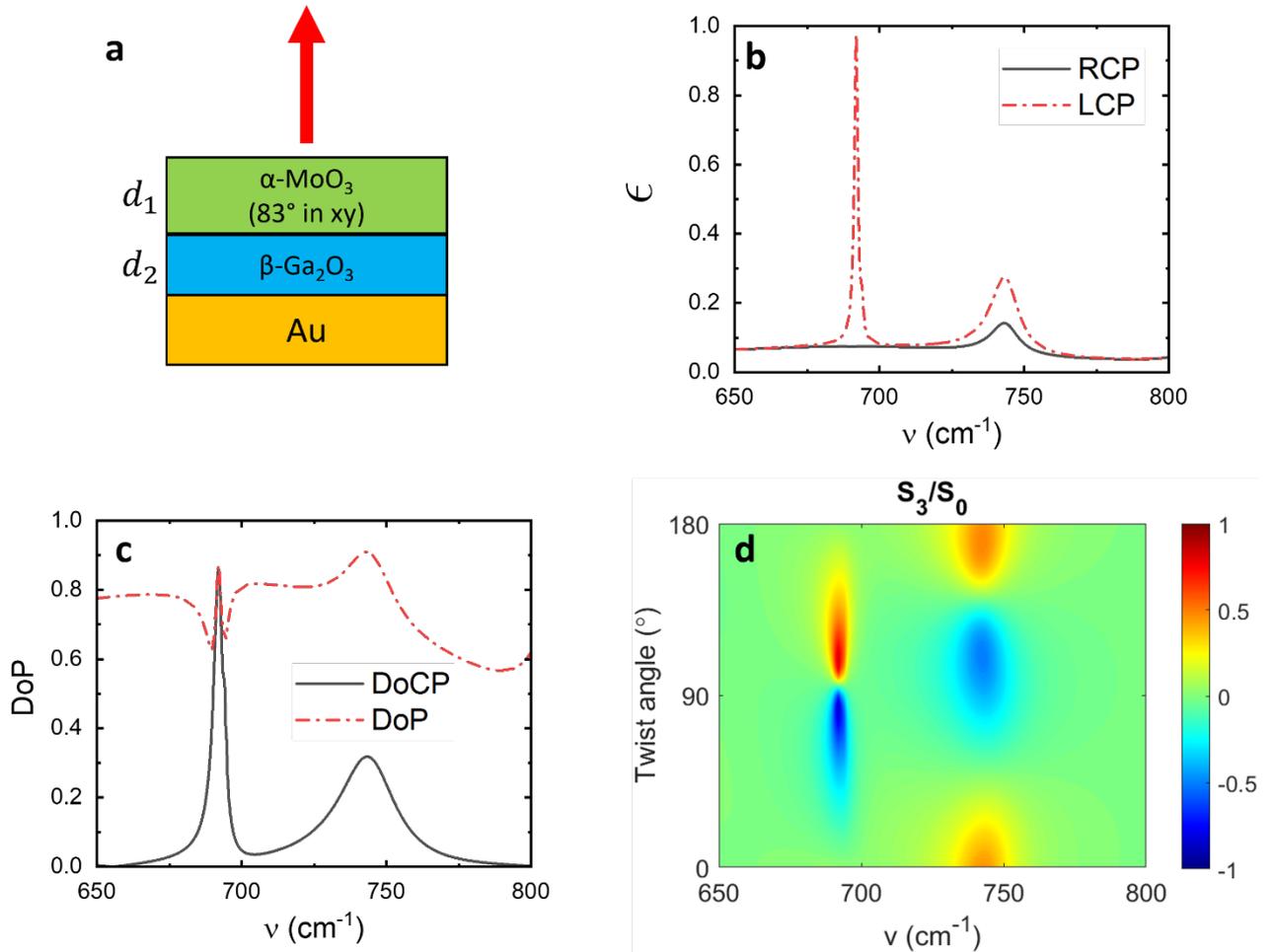

Figure 3. (a) Schematic of twisted bi-layer structure designed for ν=691 cm$^{-1}$, obtained by adding a α-MoO$_3$ layer on top of the β-Ga$_2$O$_3$ layer; (b) circularly polarized emissivity and (c) degree of polarization of the twisted bi-layer structure; (d) S3/S0 as a function of the twist angle and of the frequency. We note

that by proper tuning of the twist angle it is possible to switch from LCP to RCP emission at ν=691 cm$^{-1}$.

A similar performance is achieved for the second scheme adding a 324 nm thick layer of α-MoO$_3$ as sketched in Figure 4a, rotated in the *x-y* plane of 114° with respect to the *x'y'* axes of the β-Ga$_2$O$_3$ layer. We note that in this case, we obtain a resulting DoP = DoCP = 0.85 (figure 4c) and the thermal emitted radiation at 737 cm$^{-1}$ is mostly left circularly polarized (figure 4b). However, right-hand circular polarization at 737 cm$^{-1}$ can be obtained similarly to the previous scheme, by varying the twist angle of the MoO$_3$ layer as shown in Figure 4d.

We finally explore the robustness of the proposed schemes by varying the emission angle i.e. the Azimuth angle (ϕ) and Zenith angle (θ)) to show that the results are maintained over a large angular range for both schemes (Figure 5a and 5b respectively).

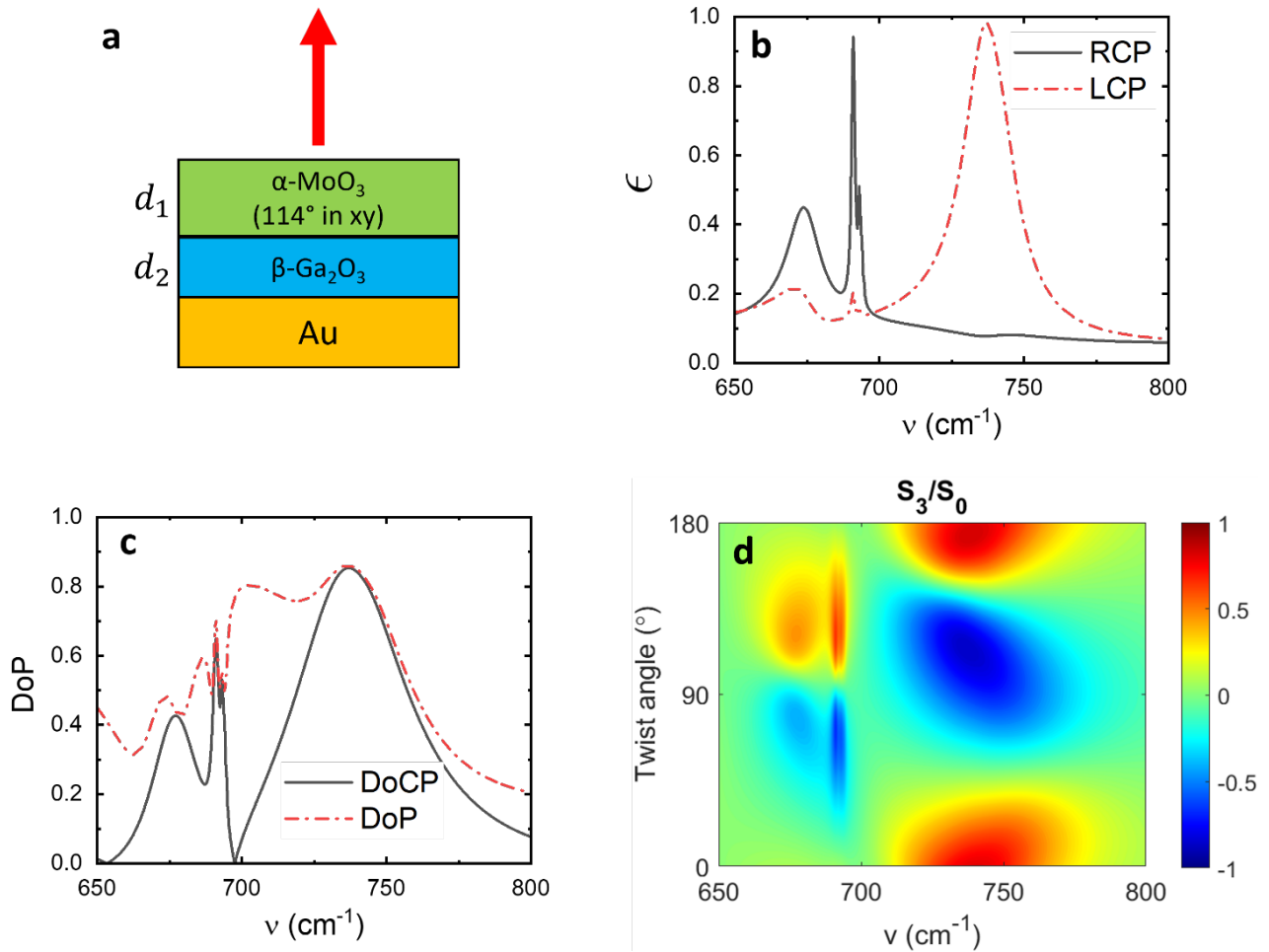

Figure 4. (a) Schematic of twisted bi-layer structure designed for ν=737 cm$^{-1}$, obtained by adding a α-MoO$_3$ layer on top of the β-Ga$_2$O$_3$ layer; (b) circularly polarized emissivity and (c) degree of polarization of the twisted bi-layer structure; (d) S3/S0 as a function of the twist angle and of the frequency. We note that by proper tuning of the twist angle it is possible to switch from LCP to RCP emission at ν=737 cm$^{-1}$.

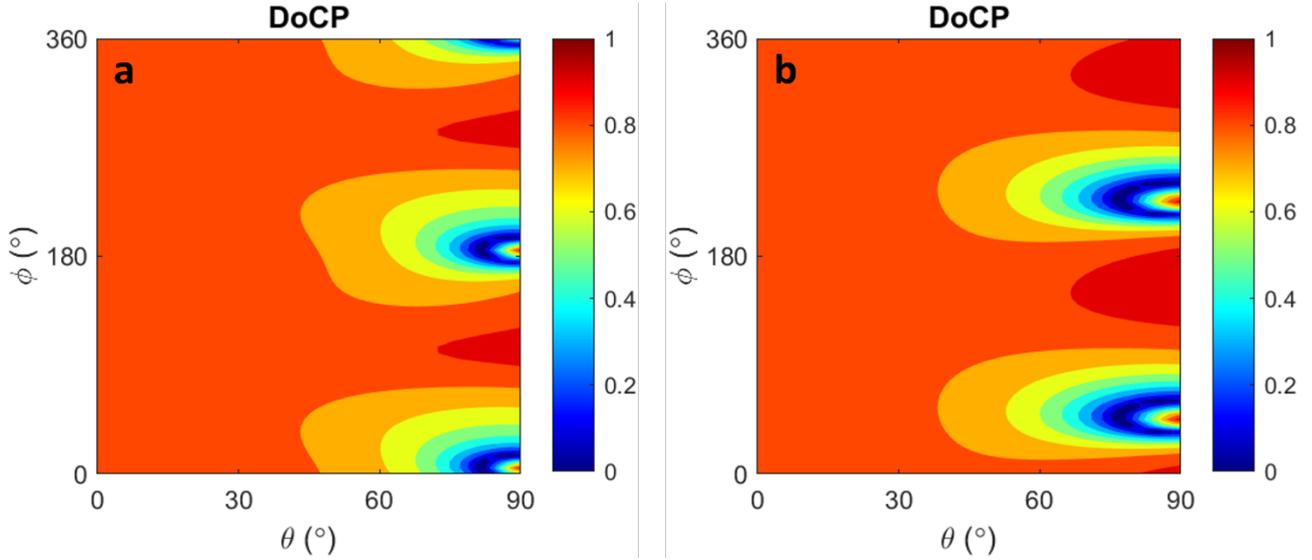

Figure 5. DoCP of thermal radiation emitted (a) at ν=691 cm$^{-1}$ for the structure sketched in figure 3a and (b) at ν=737 cm$^{-1}$ for the structure sketched in figure 4a as a function of angle (ϕ) and Zenith angle (θ).

Further evidence of the robustness of the bi-layer approach is reported in Figure 6. We performed single parameters sweep (on Ga$_2$O$_3$ layer and on MoO$_3$ layer thickness) around the optimal conditions for scheme 1 (Figure 6 a,b) and scheme 2 (Figure 6 c,d). We show that high degree of circular polarization is maintained over a large range of the thickness of the layers corresponding to hundreds of nm. We note that scheme 1 allows for a narrow band emission with a good tunability of 5 cm$^{-1}$, varying the thickness of the Ga$_2$O$_3$ layer and maintaining a DoCP higher than 0.7. On the other hand, scheme 2 emits on a larger bandwidth and it is extremely robust with respect to tolerances on the layer's thickness.

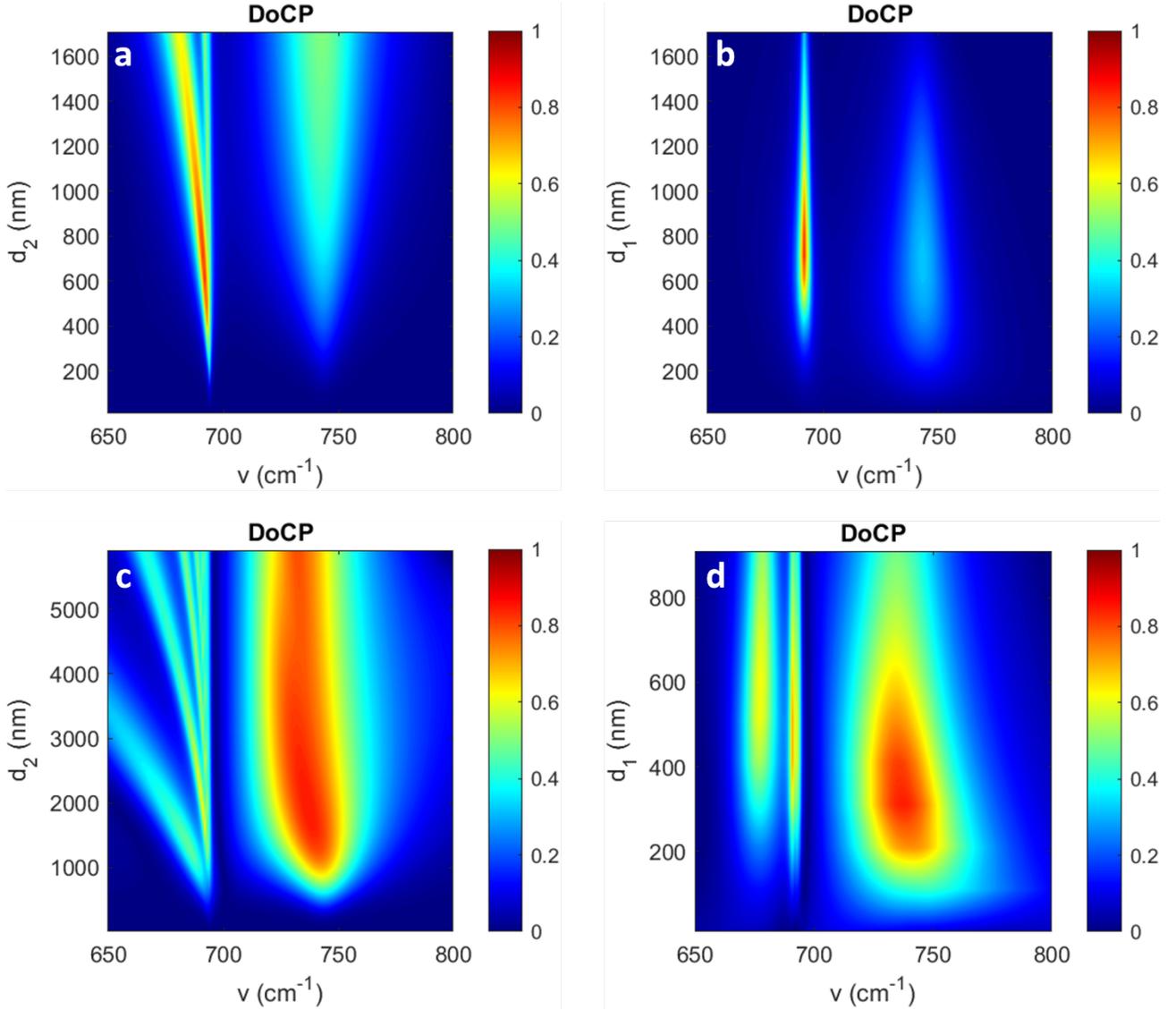

Figure 6. DoCP as a function of the frequency and the (a) β–Ga2O3 layer thickness, (b) α–MoO3 layer thickness for scheme 1. DoCP as a function of the frequency and the (c) β–Ga2O3 layer thickness, (d) α–MoO3 layer thickness for scheme 2.

## 5. CONCLUSION

We proposed a lithography-free method based on a double-twisted layer scheme to obtain circularly polarized thermal radiation in the mid-IR, at two specific frequencies related to β-Ga$_2$O$_3$ optical phonons excitation. The almost linearly polarized emission from the single slab is converted into circular polarization by adding a properly sized and tilted α-MoO$_3$ layer acting as a quarter-wavelength plate. In both cases, we take advantage of the strong natural anisotropy related to the low symmetry class of the investigated materials without the need for further processing techniques. The achieved degree of circular polarization is higher than 0.85 for the two proposed schemes and it is maintained over a broad range of angles of emission. We believe that this approach could lead to the development of low-cost, integrated thermal sources of circularly polarized light for IR sensing applications.

## ACKNOWLEDGEMENT

M.A. thanks the SAPIENZA University of Rome and the Department of Basic and Applied Sciences for Engineering for hospitality during his stay in Rome under the visiting professor program, where this work has been initiated. M.C, M.C.L, M.A. and Z.M.Z. acknowledge the KITP program 'Emerging Regimes and Implications of Quantum and Thermal Fluctuational Electrodynamics' 2022, where part of this work has been done. This research was supported in part by the National Science Foundation under Grant No. PHY-1748958. C.Y. was supported by the National Science Foundation (CBET-2029892).